\documentclass[
    ,final            
  ]
  {aipproc}

\layoutstyle{6x9}

\usepackage{epsfig}
\usepackage{amsmath}
\usepackage{mathrsfs,amssymb}
\usepackage{latexsym}
\usepackage{dsfont}
\usepackage{braket}
\usepackage{dcolumn}

\newcolumntype{d}[0]{D{.}{.}{-1}}

\newcommand{\columnspacefixer}[0]{\hspace{-4mm}\!\!\!\!}

\newcommand{\newhline}[0]{\\[1mm] \hline}

\newcommand{\superpartner}[1]{\ensuremath{{\tilde{{#1}}}}}
\newcommand{\mxone}[0]{\ensuremath{m_{{{\tilde{\chi}}^{0}_{1}}}}}
\newcommand{\xone}[0]{\ensuremath{{{\superpartner{{\chi}}}_{1}^{0}}}}

\newcommand{\Ktopi}[0]{\ensuremath{{K^{-} {\to} {\pi}^{-} {\xone} {\xone}}}}
\newcommand{\BtoK}[0]{\ensuremath{{B^{-} {\to} K^{-} {\xone} {\xone}}}}
\newcommand{\BtoKpi}[0]{\ensuremath{{B^{-} {\to} K^{-} / {\pi}^{-} {\xone} {\xone}}}}

\newcommand{\ie}[0]{\textit{i.e.}}

\def\lsim{\raise0.3ex\hbox{$\;<$\kern-0.75em\raise-1.1ex\hbox{$\sim\;$}}}
\def\gsim{\raise0.3ex\hbox{$\;>$\kern-0.75em\raise-1.1ex\hbox{$\sim\;$}}}

\begin{document}

\title{Rare meson decays into very light neutralinos}
\rightline{PITHA 09/24}







\author{Ben O'Leary}{
  address={Institut f{\"{u}}r Theoretische Physik, RWTH Aachen University, 52056 Aachen, Germany}
}

\begin{abstract}
  Results are presented for the two-body decays of mesons into light neutralinos and from the first complete calculation of the loop-induced decays of kaons to pions plus light neutralinos and of $B$ mesons to kaons plus light neutralinos.  The branching ratios are shown to be strongly suppressed within the MSSM with minimal flavor violation, and no bounds on the neutralino mass can be inferred from experimental data, \ie~a massless neutralino is allowed.
\end{abstract}

\maketitle


\section{Introduction }

  The particle data group~\cite{PDG} quotes a model-dependent lower limit on the mass of the lightest neutralino~\cite{Abdallah:2003xe} of ${\mxone} > 46$~GeV
, assuming the unification of the gaugino mass parameters $M_{1}$ and $M_{2}$ at some high energy scale.  The renormalization group evolution 
 then implies
\begin{equation}\label{eq:m_uni}
M_{1} = {\frac{5}{3}} {\frac{{{g'}^{2}}}{{g^{2}}}} M_{2} = {\frac{5}{3}} {\tan}^{2} {\theta}_{W} M_{2} {\approx} {\frac{1}{2}} M_{2}
\end{equation}
at the electroweak scale ($g$ and $g'$ denote the $SU(2)_{L}$ and $U(1)_{Y}$ gauge couplings respectively, and ${\theta}_{W}$ the weak mixing angle).  The experimental 
 bound for the lighter chargino, $m_{{{\tilde{{\chi}}}^{{\pm}}_{1}}} > 94$~GeV
~\cite{PDG}, places a lower bound on $M_{2}$ (and on the higgsino mass parameter ${\mu}$) and indirectly, through Eq.~(\ref{eq:m_uni}), on $M_{1}$.  The bounds on $M_{1}$, $M_{2}$ and ${\mu}$, as well as the lower bound on ${\tan}{\beta}$ 
from the LEP Higgs searches \cite{lhwg}, then in turn give rise to the above lower bound on ${\mxone}$.

This paper presents the results of an investigation~\cite{source_paper} into a more general MSSM scenario where Eq.~(\ref{eq:m_uni}) is not assumed and 
 $M_{1}$ and $M_{2}$ are treated as independent parameters.  It has been shown that in such a scenario one can adjust $M_{1}$ and $M_{2}$ such that the lightest neutralino is massless~\cite{Bartl:1989ms, Gogoladze:2002xp,Dreiner:2003wh}.  While a very light or massless neutralino cannot provide the cold dark matter content of the universe, it is consistent with all existing laboratory constraints and astrophysical and cosmological observations~\cite{Dreiner:2009ic}.

\section{Meson decays into light neutralinos}

\begin{figure}
\begin{tabular}{ c c c }
  \includegraphics[width=6cm]{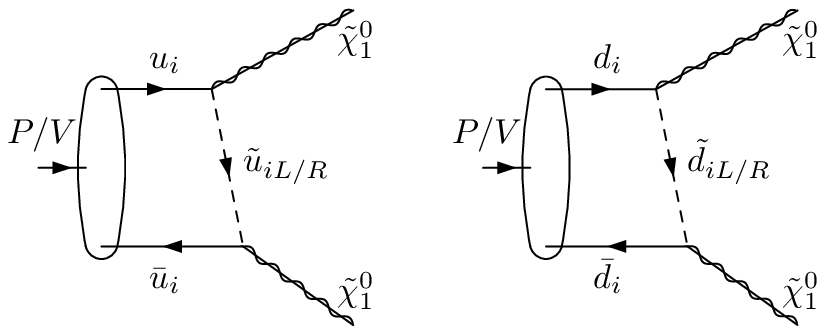} & \hspace{1cm} & \includegraphics[width=3cm]{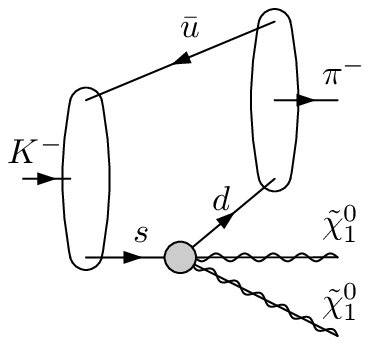}
\end{tabular}
  \caption{Feynman diagrams for the decay $P/V {\to} {\xone} {\xone}$ (left and middle) and generic overview for ${\Ktopi}$ (right), omitting diagrams with the ${\xone}$s crossed.}
  \label{fig:Feynman_diagrams}
\end{figure}

The tree-level contributions to the decays of pseudoscalar $(P)$ or vector $(V)$ mesons, $P/V {\to} {\xone} {\xone}$, are mediated by ${\superpartner{u}}_{{iL/R}}$ and ${\superpartner{d}}_{{iL/R}}$ $t$-channel exchange, Fig.~\ref{fig:Feynman_diagrams}.

\begin{figure}
\begin{tabular}{ c c }
  \includegraphics[width=5cm]{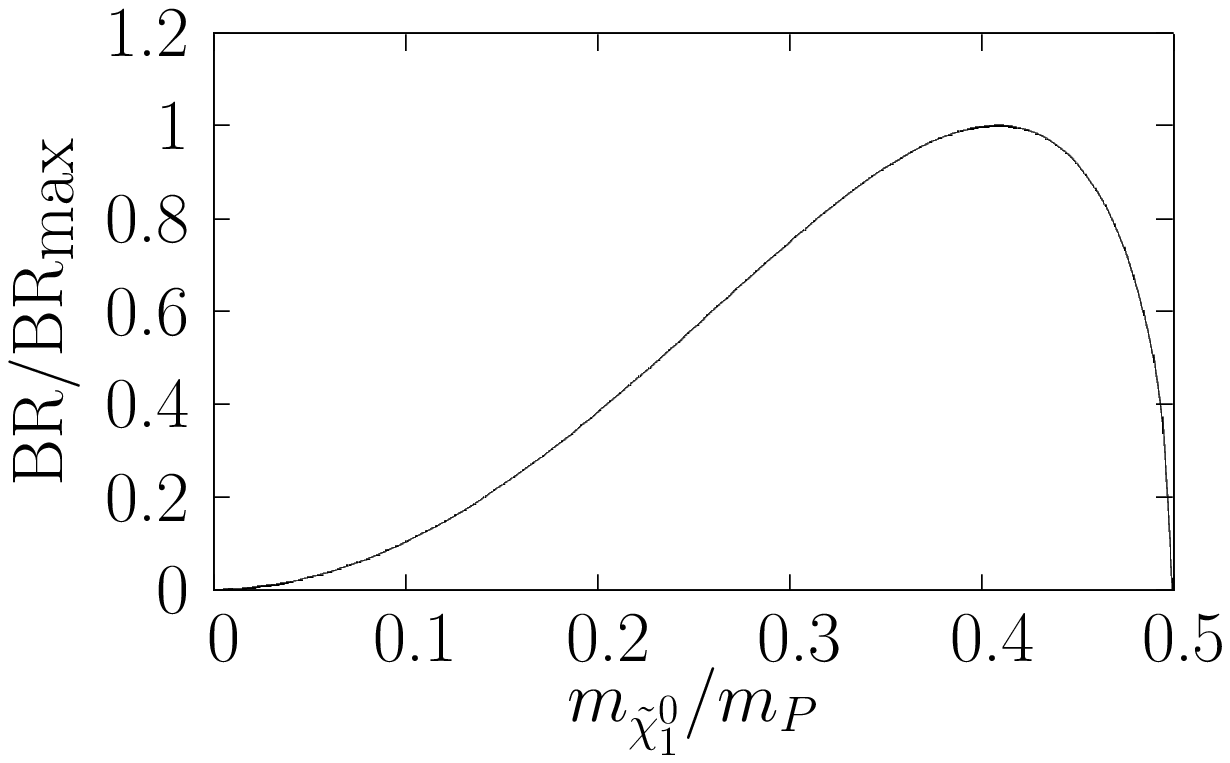} & \includegraphics[width=5cm]{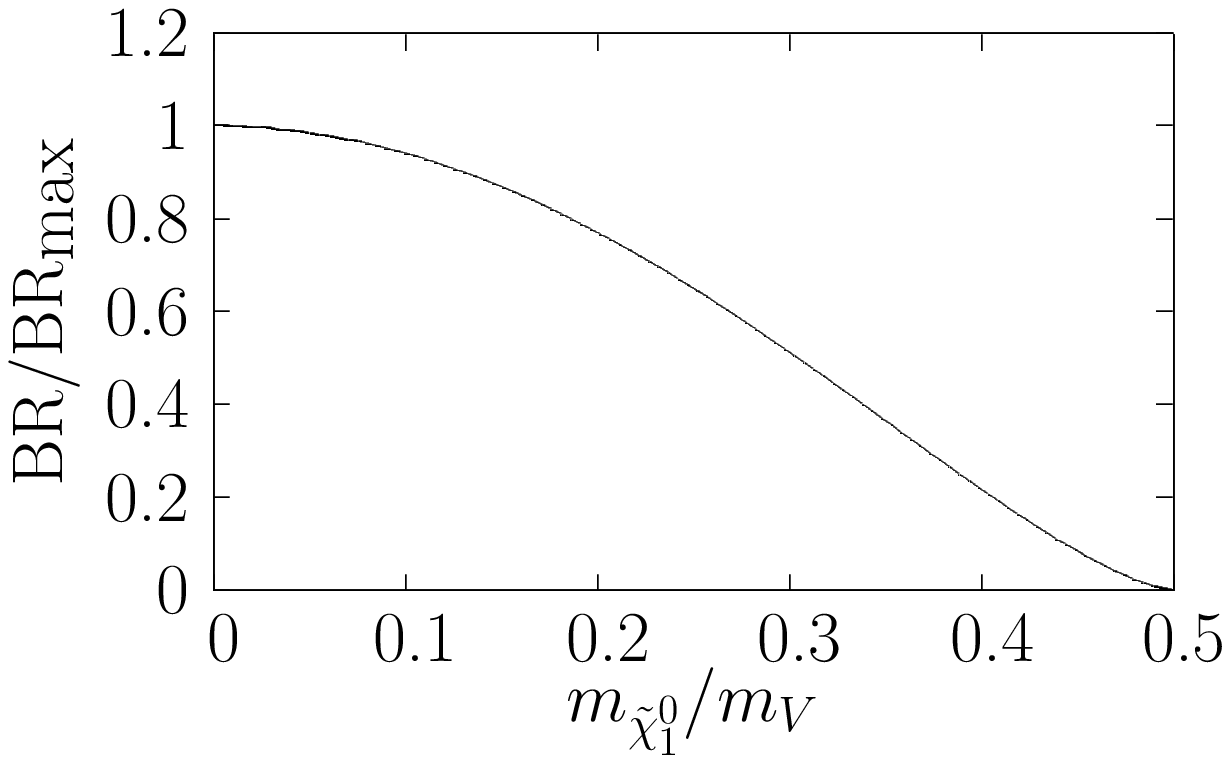}
\end{tabular}
  \caption{\label{fig:two_body_decay_mass_dependence_figures} The BRs for a pseudoscalar ($P$) and vector meson ($V$) decay $P/V {\to} {\xone} {\xone}$ against the ratio ${\mxone} / m_{P/V}$ normalized to the maximal BR.}
\end{figure}

Since the branching ratio (BR) depends on the ratio ${\mxone} / m_{P}$ in exactly the same way for each pseudoscalar meson $P$ and differs only by an overall factor, and similarly for each vector meson $V$, the dependences are plotted once, in Fig.~\ref{fig:two_body_decay_mass_dependence_figures}.

The decay of a pseudoscalar to a fermion-antifermion pair requires at least one of the fermions to be massive, due to the required helicity flip.  In Fig.~\ref{fig:two_body_decay_mass_dependence_figures} we see that the BR for pseudoscalar decays goes to zero as ${\mxone} {\to} 0$.  Hence pseudoscalar decays offer no bounds for models with massless neutralinos.  Instead, the maximal BR is reached for ${\mxone} / m_{P} = 1 / {\sqrt{6}} \approx 0.41$.  Note that there is no helicity suppression for vector decays, and so increasing ${\mxone}$ only reduces the available phase space and thus the BR.  Hence, the maximal BR is reached for ${\mxone = 0}$.  Since there is interference between the contributions of different flavors of valence quark, the results presented in Tab.~\ref{tab:two_body_results_table} are for the extreme cases where the BR is maximized by decoupling squarks in such a way as to minimize destructive interference.


\begin{table}
\setlength{\extrarowheight}{1mm}
\begin{tabular}{c d@{${\columnspacefixer}$}l d@{${\columnspacefixer}{\;}$}l c d@{${\columnspacefixer}$}l d@{${\columnspacefixer}{\;}$}l}
\hline
Meson & \multicolumn{2}{c}{BR${}_{max}$(${\to} {\xone} {\xone}$)} & \multicolumn{2}{c}{Exp.~bound} & Meson & \multicolumn{2}{c}{BR${}_{max}$(${\to} {\xone} {\xone}$)} & \multicolumn{2}{c}{Exp.~bound}
\newhline
${\pi}^{0}$ & $1.63$ & ${\times} 10^{{-10}}$ & $2.7$ & ${\times} 10^{{-7}}$ & ${\rho}^{0}$ & $8.01$ & ${\times} 10^{{-15}}$ & \multicolumn{2}{c}{none}
\newhline
${\eta}$ & $7.60$ & ${\times} 10^{{-11}}$ & $6$ & ${\times} 10^{{-4}}$ & ${\omega}$ & $7.51$ & ${\times} 10^{{-14}}$ & \multicolumn{2}{c}{none}
\newhline
${{\eta}'}$ & $3.83$ & ${\times} 10^{{-12}}$ & $1.4$ & ${\times} 10^{{-3}}$ & ${\phi}$ & $1.57$ & ${\times} 10^{{-13}}$ & \multicolumn{2}{c}{none}
\newhline
$J/{\psi}$ & $5.12$ & ${\times} 10^{{-9}}$ & $5.9$ & ${\times} 10^{{-4}}$ & ${\Upsilon}$ & $4.47$ & ${\times} 10^{{-8}}$ & $2.5$ & ${\times} 10^{{-3}}$
\newhline
\end{tabular}
\caption{\label{tab:two_body_results_table} Maximized BRs for two-body decays of pseudoscalar mesons to neutralinos, assuming minimal destructive interference and all non-decoupled squarks having mass of $100$~GeV.}
\end{table}

\section{${\bf{\Ktopi}}$ and ${\bf{\BtoK}}$}
\label{MFV_section}

There is no required helicity flip for pseudoscalar decay into three-body final states, and so there could be significant bounds from the decays ${\Ktopi}$ and ${\BtoK}$.

The flavor-changing vertex at the level of the valence quarks in the mesons can be represented by a set of four-fermion operators, either from integrating out squarks with mass eigenstates that are not aligned with the quark mass eigenstates, or from loops involving charged flavor-changing currents.


If one requires Minimal Flavor Violation (MFV), such that the Cabibbo-Kobayashi-Maskawa (CKM) matrix is the only source of flavor-violation, then these decays happen only through flavor-changing loops.  Explicit Feynman diagrams for these loops can be found in Ref.~\cite{source_paper}, as can the details of the calculation.  Quoted here are the BRs obtained when the sparticle mass spectrum is taken to be that of the SPS points 1a, 2, 3, 4 and 5, with $M_{1}$ adjusted such that ${\mxone} = 0$ (denoted ``pseudo-SPS'' points).

\begin{table}
\begin{tabular}{l d@{${\columnspacefixer}$}l d@{${\columnspacefixer}$}l d@{${\columnspacefixer}$}l d@{${\columnspacefixer}$}l d@{${\columnspacefixer}$}l d@{${\columnspacefixer}$}l}
\hline
pseudo-SPS & \multicolumn{2}{c}{1a} & \multicolumn{2}{c}{2} & \multicolumn{2}{c}{3} & \multicolumn{2}{c}{4} & \multicolumn{2}{c}{5}
\newhline \\[-5mm] \hline\\[-4mm]
BR${}_{{\Ktopi}}$ & $3.28$ & ${\times} 10^{-16}$ & $1.47$ & ${\times} 10^{-18}$ & $6.99$ & ${\times} 10^{-17}$ & $8.76$ & ${\times} 10^{-17}$ & $5.12$ & ${\times} 10^{-16}$
\newhline
BR/Exp. & $1.90$ & ${\times} 10^{-6}$ & $8.49$ & ${\times} 10^{-9}$ & $4.04$ & ${\times} 10^{-7}$ & $5.06$ & ${\times} 10^{-7}$ & $2.96$ & ${\times} 10^{-6}$
\newhline \\[-5mm] \hline\\[-4mm]
BR${}_{{\BtoK}}$ & $3.35$ & ${\times} 10^{-10}$ & $2.48$ & ${\times} 10^{-12}$ & $7.19$ & ${\times} 10^{-11}$ & $2.53$ & ${\times} 10^{-10}$ & $7.14$ & ${\times} 10^{-10}$
\newhline
BR/Exp. & $2.39$ & ${\times} 10^{-5}$ & $1.77$ & ${\times} 10^{-7}$ & $5.14$ & ${\times} 10^{-6}$ & $1.81$ & ${\times} 10^{-5}$ & $5.10$ & ${\times} 10^{-5}$
\newhline
\end{tabular}
\caption{Numerical values for calculated BRs for ${\Ktopi}$ and ${\BtoK}$ at the various pseudo-SPS points described in the text.  The rows denoted BR/Exp. show the ratios of the calculated BR to the experimental value of the BR for $K^{-} {\to} {\pi}^{-} {\nu} {\bar{{\nu}}}$ ($1.73 {\times} 10^{-10}$) or to the experimental upper bound for $B^{-} {\to} K^{-} {\nu} {\bar{{\nu}}}$ ($1.4 {\times} 10^{-5}$).}
\label{Ktopi_numerical_table}
\end{table}

\begin{figure}
  \centering
  \begin{minipage}[b]{6cm}
     \includegraphics[scale=0.7]{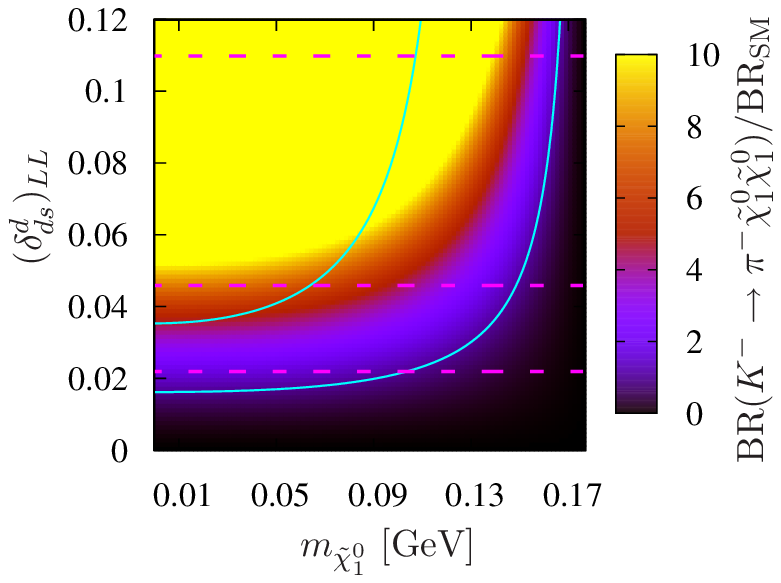}  
  \end{minipage}
  \begin{minipage}[b]{6cm}
     \includegraphics[scale=0.7]{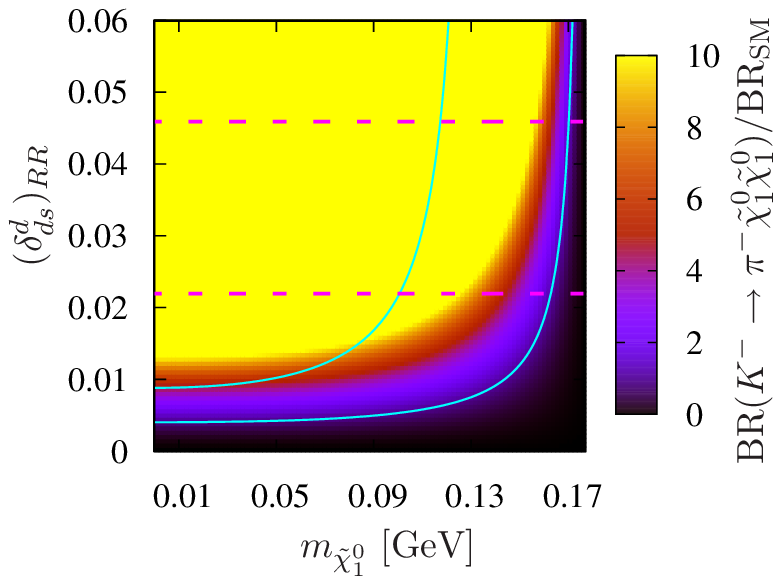}
  \end{minipage}
  \caption{\label{K_to_Pi_LL} BRs for ${\Ktopi}$ for varying ${\mxone}$ and $( {\delta}^{d}_{ds} )_{LL}$ (left) and $( {\delta}^{d}_{ds} )_{RR}$ (right), normalized to the SM prediction for $K^{-} {\rightarrow} {\pi}^{-} {\nu} {\bar{{\nu}}}$ ($8.5 {\times} 10^{-11}$~\cite{SM_prediction}).  The lightest grey (yellow) region corresponds to normalized BRs ${\geq} 10$.  The upper (lower) solid gray (turquoise) line corresponds to the experimentally measured BR $+ 2 {\sigma}$ for $K^{-} {\rightarrow} {\pi}^{-} {\nu} {\bar{{\nu}}}$, multiplied by a correction factor (see Ref.~\cite{source_paper} for details).  The dashed lines show upper bounds on the mass insertions $(\delta^d_{ds})_{LL/RR}$ obtained from $K^{0}$--${\bar{{K^{0}}}}$ mixing~\cite{Ciuchini:1998ix} for different ratios of squark and gluino masses.  \vspace{-5mm}}
\end{figure}

If one relaxes the constraint of MFV, these three-body decays can happen at tree-level.  In Fig.~\ref{K_to_Pi_LL} the BRs for ${\Ktopi}$ are presented for varying ${\mxone}$ and mass insertion parameter $( {\delta}^{d}_{ij} )_{XY} {\equiv} ( {\delta}m^{2}_{d,XY} )_{ij} / {\tilde{m}}^{2}$, where ${\delta}m^{2}_{q,XY}$ are the off-diagonal elements of the squark mass-squared matrix in the super-CKM basis and ${\tilde{m}}^{2}$ is the ``average'' squark mass squared, which is assumed to be $500$ GeV.  Results for the cases ${\BtoKpi}$ can be found in Ref.~\cite{source_paper}.

\section{Conclusion}
\label{sec:conclusion}

It has been shown that the supersymmetric BRs in various MSSM scenarios are several orders of magnitude smaller then the SM processes with neutrinos instead of neutralinos in the final state.  Consequently, no bounds on the neutralino mass can be inferred from rare meson decays in the MSSM with minimal flavor violation.  However, the BRs for the ${\Ktopi}$ and ${\BtoK}$ decays may be significantly enhanced when one allows for non-minimal flavor violation, and new constraints on the MSSM parameter space for such scenarios have been presented.


\begin{theacknowledgments}
The work presented was supported in part by the European Community's Marie-Curie Research Training Network under contract MRTN-CT-2006-035505 ``Tools and Precision Calculations for Physics Discoveries at Colliders'', the DFG SFB/TR9 ``Computational Particle Physics'' and the Helmholtz Alliance ``Physics at the Terascale''.
\end{theacknowledgments}

\end{document}